\documentclass[unnumsec,webpdf,contemporary,large]{oup-authoring-template}

\usepackage[utf8]{inputenc}
\usepackage{acronym}
\usepackage{graphicx}
\usepackage{hyperref}
\usepackage[english]{babel}

\begin{document}

\journaltitle{Microscopy and Microanalysis}

\copyrightyear{2023}

\appnotes{Paper}

\firstpage{1}

\title{Calibrating coordinate system alignment in a scanning transmission electron microscope using a digital twin}

\author[1, $\ast$]{Dieter Weber\ORCID{0000-0001-6635-9567}}
\author[2, 3]{David Landers\ORCID{0000-0002-3295-9473}}
\author[4]{Chen Huang\ORCID{0000-0002-7864-8427}}
\author[4]{Emanuela Liberti\ORCID{0000-0001-5306-7936}}
\author[5]{Emiliya Poghosyan\ORCID{0000-0002-0961-731X}}
\author[3]{Matthew Bryan\ORCID{0000-0001-9134-384X}}
\author[1]{Alexander Clausen\ORCID{0000-0002-9555-7455}}
\author[6]{Daniel G.\ Stroppa\ORCID{0000-0002-7711-1839}}
\author[4, 7]{Angus I.\ Kirkland\ORCID{0000-0001-8068-1990}}
\author[5]{Elisabeth Müller\ORCID{0000-0001-8239-7109}}
\author[2]{Andrew Stewart\ORCID{0000-0002-3081-5644}}
\author[1]{Rafal E.\ Dunin-Borkowski\ORCID{0000-0001-8082-0647}}

\address[1]{\orgdiv{Ernst Ruska-Centre for Microscopy and Spectroscopy with Electrons}, \orgname{Forschungszentrum J\"ulich}, \orgaddress{\postcode{52425 J\"ulich}, \country{Germany}}}
\address[2]{\orgdiv{Department of Chemistry}, \orgname{University College London}, \orgaddress{\street{20 Gordon Street}, \postcode{London, WC1H 0AJ}, \country{UK}}}
\address[3]{\orgdiv{LETI}, \orgname{CEA}, \orgaddress{\street{17 Av.\ des Martyrs}, \postcode{38054 Grenoble}, \country{France}}}
\address[4]{\orgdiv{Rosalind Franklin Institute}, \orgname{Harwell Science and Innovation Campus}, \orgaddress{\postcode{Didcot, OX11 0QX}, \country{UK}}}
\address[5]{\orgdiv{Electron Microscopy Facility at PSI}, \orgname{Paul Scherrer Institute}, \orgaddress{\postcode{5232 Villigen}, \country{Switzerland}}}
\address[6]{\orgname{DECTRIS AG}, \orgaddress{\street{Taefernweg 1}, \postcode{5405 Baden-Daettwil}, \country{Switzerland}}}
\address[7]{\orgdiv{Department of Materials}, \orgname{University of Oxford}, \orgaddress{\postcode{Oxford, OX1 3PH}, \country{UK}}}

\corresp[$\ast$]{Corresponding author. \href{email:d.weber@fz-juelich.de}{d.weber@fz-juelich.de}}

\abstract{In four-dimensional scanning transmission electron microscopy (4D STEM) a focused beam is scanned over a specimen and a diffraction pattern is recorded at each position using a pixelated detector. During the experiment, it must be ensured that the scan coordinate system of the beam is correctly calibrated relative to the detector coordinate system. Various simplified and approximate models are used implicitly and explicitly for understanding and analyzing the recorded data, requiring translation between the physical reality of the instrument and the abstractions used in data interpretation. Here, we introduce a calibration method where interactive live data processing in combination with a digital twin is used to match a set of models and their parameters with the action of a real-world instrument.}


\maketitle

\section{Introduction}

\begin{figure}
    \centering
    \includegraphics[width=230pt]{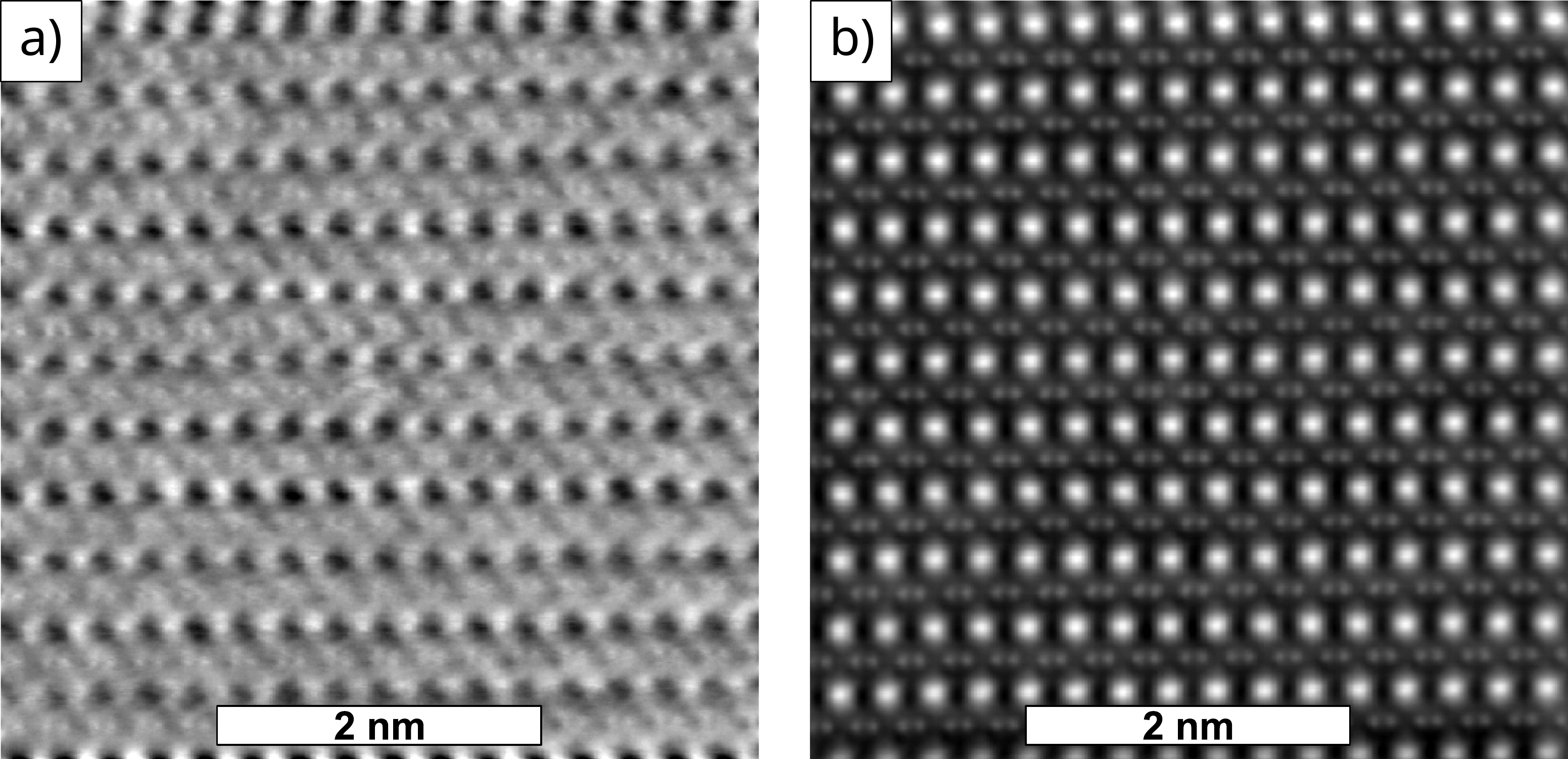}
    \caption{Impact of the rotation calibration on an integrated \ac{CoM}~\citep{Lazic2017} reconstruction of a 4D \ac{STEM} dataset of $\text{SmB}_6$, recorded with a DECTRIS ARINA on a Thermo Fisher Spectra 200: a) The calibration of the scan rotation is off by 90°. b) Correct calibration of the scan rotation.
    }\label{fig:icom}
\end{figure}

For several 4D \ac{STEM} data analysis methods such as \ac{DPC}~\citep{Hamilton1984, Lazic2016} or \ac{CoM}~\citep{Lazic2017}, strain~\citep{Usuda2005,Ozdol2015} and orientation~\citep{VILADOT2013,Rollett2014} mapping, and ptychography~\citep{Hoppe1969,Yang2017} it is essential that the relative orientation between the scan coordinate system and the detector coordinate system is known so that directions on the detector can be put into correspondence with directions in the scan coordinate system~\citep{Ning2022}. Figure~\ref{fig:icom} shows the impact of an incorrect rotation calibration on integrated \ac{CoM} reconstruction for illustration. This calibration is surprisingly difficult and error-prone in practice, and it was often not required in \ac{STEM} previously, since conventional detectors such as those used for bright field and annular dark field \ac{STEM} are rotationally symmetric. Consequently, there has been substantial recent work in both theory and practical implementation to calibrate a microscope's coordinate system as well as validate the calibration of a given 4D \ac{STEM} dataset.

Three methods are established to perform this alignment. First, the deflection distribution around atom columns in atomic-resolution \ac{STEM} can be analyzed. Since the deflection is, in a thin specimen approximation, proportional to the projected local field, a beam of negative electrons should be attracted to the positively charged atom columns~\citep{Shibata2012}. Furthermore, an electrostatic field should be free of curl. The coordinate system alignment that minimizes curl and ensures a negative divergence at atom positions is assumed to be the correct one for this method~\citep{Savitzky2021}.

Second, the self-consistency of the data in a ptychography reconstruction can be analyzed, such as described in \citep{Ning2022}.

In practice, however, real-world specimens are usually not thin enough to assume a simple direct dependence between displacement and field without multiple scattering. In particular, slight misalignments between the atom column axis and the beam axis as well as optical aberrations might lead to apparent distortions~\citep{Buerger2020}. Furthermore, atom columns are only reliably resolved if the instrument aberrations are very low and a high spatial resolution is achieved. Not every scanning transmission electron microscope is capable of this, and it is not always desirable to tune and operate the microscope in this mode. Additionally, methods based on analyzing field distribution around atom columns are only applicable to thin crystalline specimens.

The third and more conventional approach relies on over- or underfocusing the beam so that the crossover above resp.\ below the specimen acts as a point source. In this configuration, a shadow image of the specimen is projected onto the detector. If this shadow image is compared with a \ac{STEM} image, the transformation between \ac{STEM} image coordinates and shadow image, i.e.\ detector coordinates, can be determined. Instead of a separate \ac{STEM} image, a virtual detector image can be generated from a 4D \ac{STEM} dataset using a very small virtual detector, for example a single detector pixel, so that it has a high depth of field. In that case both detector shadow image and \ac{STEM} image are taken from the same 4D \ac{STEM} dataset~\citep{Hu2023}. Underfocus can be used instead of overfocus, which leads to an inverted detector image compared to overfocus.

This method relies on finding a field of view on the specimen without mirror or rotational symmetry and adjusting the microscope so that it is clearly recognizable, including exposing with sufficient dose. However, the matching procedure can be error-prone if performed manually without the help of a dedicated software stack. As an example, the direction of rotation is easily reversed if the procedure is performed manually, an additional handedness change could be missed, and different software or hardware uses different coordinate system conventions. If conventional \ac{STEM} and 4D \ac{STEM} use a different acquisition software and/or scan generator on the same microscope, which is often the case, the coordinate system of a conventional \ac{STEM} image might differ from the coordinate system of a 4D \ac{STEM} dataset.

Furthermore, different software might interpret and display the data differently. As an example, some software displays the first pixel of a detector frame in the lower left corner with the y axis pointing upwards, while other image processing software shows the first pixel in the upper left with the y axis pointing downwards. That may introduce inadvertent errors if data is inspected visually with one software to calibrate the alignment, but then processed numerically with another.

To eliminate such error sources, the calibration should be performed with the same acquisition system and software as the actual acquisition to ensure consistency. Furthermore, the calibration method and analysis methods should be validated together to interpret data and parameters in a consistent way.

Here we present a method that is derived from the defocus method. It uses automated data processing and a digital twin of the microscope to superimpose all shadow images in an over- or underfocused 4D \ac{STEM} dataset. If the transformation by the digital twin matches the actual transformation by the microscope, a sharp image is obtained. In case of a mismatch between microscope and twin, the images are not superimposed correctly and the result is blurred. This blurring is independent of symmetries in the specimen, since the translation of the projected images on the detector caused by scanning the beam breaks any symmetry.

\section{Methods}

The shadow imaging in the overfocused state is modeled with a linear ray tracing approximation. All rays are assumed to originate from a focus point above the specimen, pass through the specimen, and then hit the detector plane. Underfocus, where the focus point is below the specimen, is represented with a negative overfocus value that leads to the correct result without separate adaptation of this model. Scanning the beam shifts the position of the focus point laterally relative to the specimen, but doesn't shift the beam position on the detector or lead to a beam tilt in the specimen plane. The specimen is assumed to be thin and oriented orthogonal to the optical axis, i.e.\ not tilted or warped. All coordinate systems are right-handed by default.

\begin{figure}
    \centering
    \includegraphics[width=230pt]{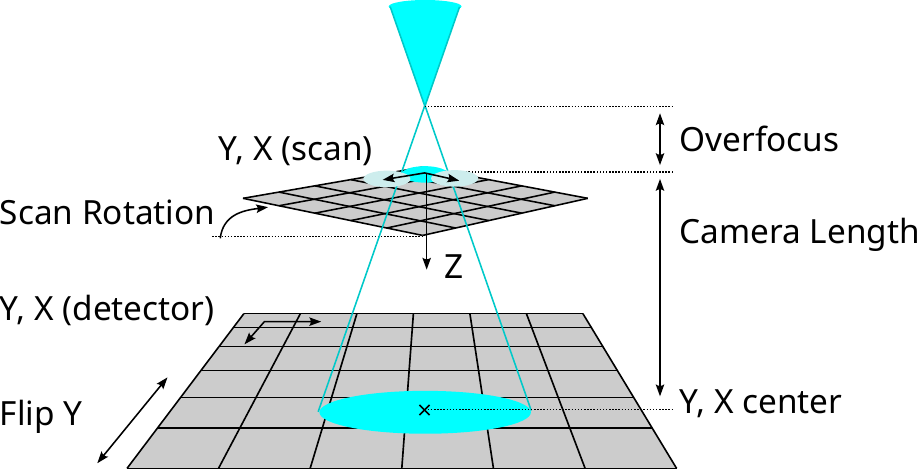}
    \caption{Abstract model of the projection with overfocused \ac{STEM} with the parameters and coordinate axes labeled.}\label{fig:abstraction}
\end{figure}

\begin{figure}
    \centering
    \includegraphics[width=230pt]{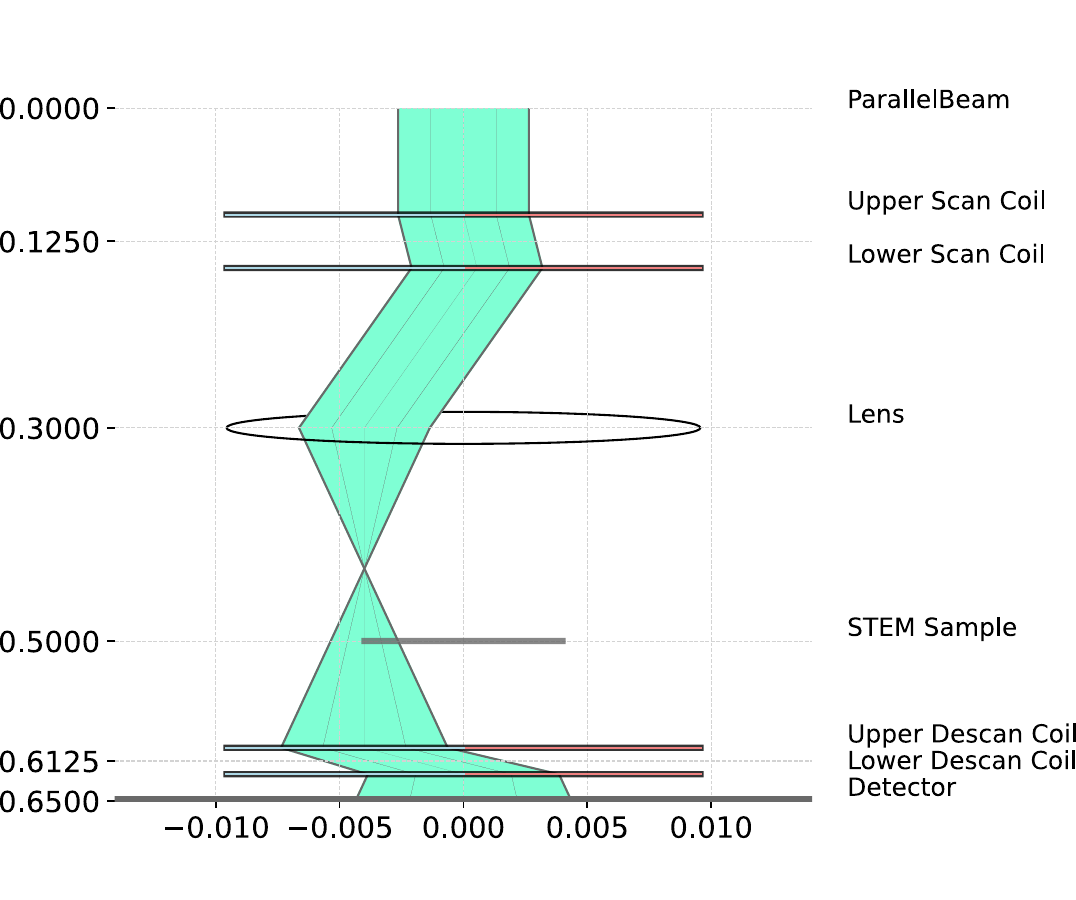}
    \caption{Visualization of the ray paths in the digital twin. The figure is generated using TEMGYM's visualization routine from the model used in the calculations with the actual dataset (Figures~\ref{fig:miscalibrated}-\ref{fig:numerical}). The overfocus and the scan step are exaggerated by a factor of $10^4$ to make them visible. Note how this model is closer to the physical reality of a scanning transmission electron microscope than the abstract model in Figure~\ref{fig:abstraction}. TEMGYM Basic translates the parameters of the abstract model into parameters for simulated optical elements such as deflectors and lenses to yield results equivalent to the abstract model.}\label{fig:visualization}
\end{figure}

With these assumptions in place, the electron-optical setup of this imaging mode is described with the following parameters:
\begin{itemize}
    \item \textbf{Overfocus} Virtual vertical distance of the beam focus over the specimen plane under the assumption of straight propagation between focus point and specimen. Underfocus is specified as a negative overfocus value.
    \item \textbf{Scan Pixel Size} Lateral distance between scan points. The scan is assumed to be rectangular with uniform step size in both directions.
    \item \textbf{Camera Length} Virtual vertical distance between specimen and detector under the assumption of straight ray propagation between specimen and detector. In real microscopes the electrons pass through a set of projection lenses that allow adjusting this virtual distance.
    \item \textbf{Detector Pixel Size} Size resp.\ pitch of detector pixels. The detector is assumed to be rectangular with equidistant square pixels without gaps.
    \item \textbf{Y and X Center} Pixel position on the detector that is defined as the beam center, i.e. where the ray from the focus point straight down orthogonal through the specimen plane hits the detector.
    \item \textbf{Scan Rotation} Rotation between the scan coordinate system in scan pixel units and the detector pixel coordinate system in pixel units. Physically, this is composed of three components: Actual rotation of the scan, rotation imposed by the projection system, and rotation of the detector. In the simple model used here these three rotations are cumulative and can be described with a single parameter.
    \item \textbf{Flip Y} Invert the detector row index (y) axis. This encodes a handedness change between detector and scan. Any combination of axis inversion and rotation for both detector and scan can be described with a combination of Scan Rotation and Flip Y.
\end{itemize}

Figure~\ref{fig:abstraction} shows a schematic of this model of the experiment with labels for the parameters.

The parameters are chosen to correspond with the usual parameters that are available from a microscope's user interface. However, it is not guaranteed that they are also interpreted in the same way. As an example, for Thermo Fisher microscopes the parameter ``Scan Rotation'' describes a clockwise rotation of the \ac{STEM} image on the screen, while our implementation interprets it as a clockwise rotation of the scan coordinate system around the beam (Z) axis, following the convention~\citep{LiberTEM-coordinates} in the LiberTEM software~\citep{Clausen2023}. That means the parameter ``Scan Rotation'' rotates in opposite direction between the Thermo Fisher software and LiberTEM.

``Flip Y'' was chosen to describe a handedness difference since it makes it easy to account for an inverted Y axis of a detector.

The parameters Overfocus $o$, Scan Pixel Size $s$, Camera Length $l$ and Detector Pixel Size $d$ are not independent, but their ratios define a single scaling factor $M = \frac{s l}{d o}$ between scan pixel coordinates and detector pixel coordinates. Since Scan Pixel Size, Camera Length and Detector Pixel Size are typically calibrated by other means or are fixed physical distances, Overfocus is a natural choice to adjust the scaling factor.

TEMGYM Basic~\citep{Landers2023} was used to construct a digital twin for a \ac{STEM} instrument (Figure~\ref{fig:visualization}) that can accept these parameters and interprets them in a consistent way to match the abstraction shown in Figure~\ref{fig:abstraction}. This digital twin can calculate both the detector pixel position and the specimen pixel position of rays as a function of diffraction angle and scan position. With this information, the shadow image projected onto the detector can be transformed into images in scan pixel coordinates. This model assumes that changing the focus doesn't affect rotation and handedness.

\begin{figure}
    \centering
    \includegraphics[width=230pt]{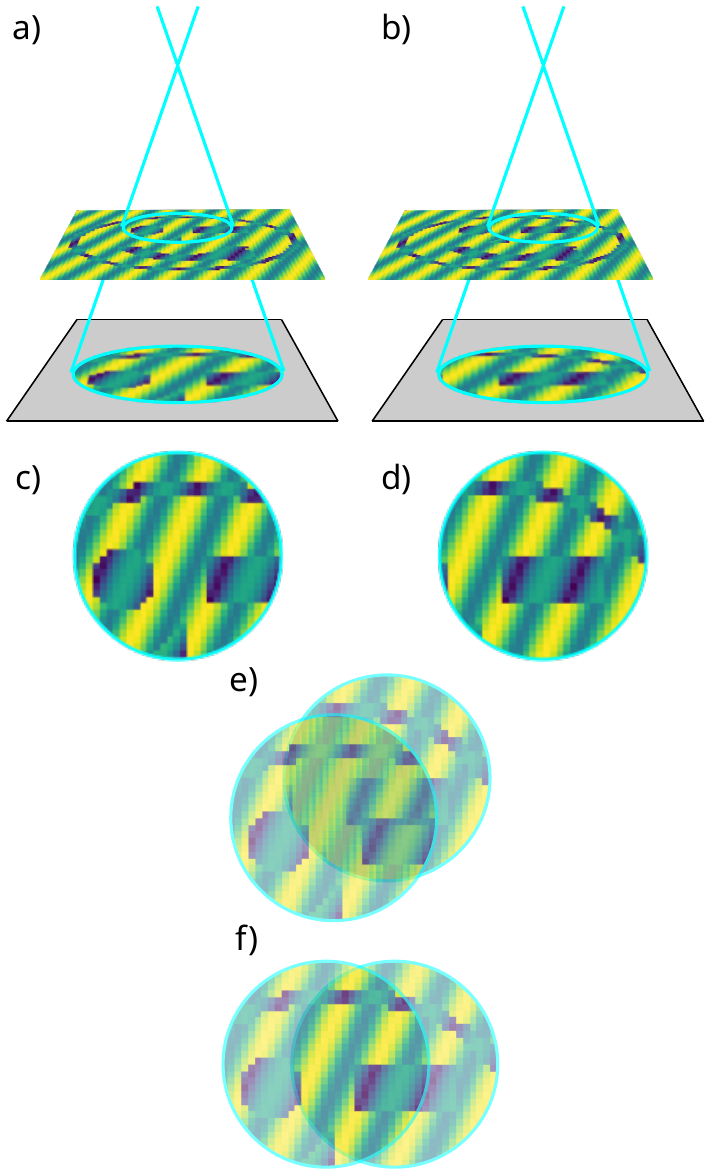}
    \caption{Illustration of the superposition of projected shadow images: a) Projection at first scan position. b) Projection at second scan position. c) Projected image from a). d) Projected image from b). e) Incorrect superposition of the projected images: The features don't align. This will lead to blurring if many images from different scan positions are superimposed incorrectly. f) Correct superposition: The features align and the superposition of many images from different scan positions creates a sharp image of the object.}\label{fig:example}
\end{figure}

If the digital twin describes the action of the microscope correctly, superimposing the images from an overfocused 4D \ac{STEM} experiment after transforming into scan pixel coordinates will form a sharp image of the specimen (Figure~\ref{fig:example}). This superimposed image will also correspond to a virtual \ac{STEM} detector image formed from a single detector pixel, for example the central ray like described in~\citep{Hu2023}. However, if the transformation of the digital twin differs from the actual microscope, the images will not be superimposed correctly, leading to a blurred result.

By adjusting the parameters of the digital twin while continuously performing the superposition operation, a user or algorithm can optimize the model parameters until the result is as sharp as possible. A possible metric for sharpness resp.\ blurriness was described by~\cite{Crete2007}. The process of optimizing sharpness is similar to conventional focusing and two-fold astigmatism adjustment, meaning it is intuitive for microscope users.

The data processing routine is implemented as a \ac{UDF} called ``OverfocusUDF'' that performs the superposition and extracts the additional control images. This allows for fast offline and live processing together with live visualization and dynamic parameter updates.

For a smooth user interaction the superposition should be performed very quickly. TEMGYM Basic does trace rays at high speed, but it is still too slow to determine the target scan coordinates for each data point in a 4D \ac{STEM} dataset in real time. For that reason a linear transformation from detector pixel position and scan position to specimen pixel position is derived from a set of sample rays using a least square optimization. This transformation is applied using an optimized kernel written in Numba.

In many cases the initial parameters will be far off their ideal values, resulting in a featureless superposition image. Since multiple parameters have to match at least approximately in order to form an image with recognizable features, two helper tranformations were implemented:

By selecting a single detector pixel and plotting its value as a function of scan position, a virtual detector image with very large depth of focus can be obtained, as described in~\citep{Hu2023} (Figures~\ref{fig:miscalibrated} - \ref{fig:fine}, label ``OverfocusUDF: point''). As long as the dose and detector resolution is sufficient, this will be a recognizable image of the specimen in scan coordinates. Second, a single shadow image on the detector can be transformed to scan coordinates without superimposing it with shadow images from other scan positions (Figures~\ref{fig:miscalibrated} - \ref{fig:fine} label ``OverfocusUDF: selected'').

If the parameters are adjusted with sufficient accuracy (Figures~\ref{fig:fine} and~\ref{fig:numerical}), these two images will look the same. By visually comparing them, a user can roughly adjust the overfocus value by comparing the scale, and the rotation and handedness by observing a unique feature on the specimen (Figures \ref{fig:miscalibrated} -- \ref{fig:rough}). This coarse adjustment will usually be accurate enough to observe features in the superimposed image (Figure~\ref{fig:rough}), allowing for further incremental optimization of all parameters following a protocol similar to optimizing focus and astigmatism. 

The model currently doesn't account for descan error. However, the presence of descan error is checked by summing up detector images without transformation. Without descan error, the beam should have the same position on the detector independent of the scan position. That means that a sum of all detector images should yield a sharp image of the aperture that forms the primary beam.

Additionally, a loss function suitable for numerical optimization can be derived from the calculation of the blurriness of the superimposed image as a function of Scan Rotation and Overfocus. A utility function is included with the ``OverfocusUDF'' that creates such a loss function by wrapping other parameters for executing it, such as LiberTEM Context, dataset and the remaining model parameters, in a closure~\citep{Sussman1998}. The returned function accepts only the Scan Rotation and Overfocus as a single vector of two values, and maps them from a range of [(-10, 10), (-10, 10)] to a sensible interval of $\pm 10 \deg$ resp. $\pm 25 \%$ around their starting value for conditioning. This ensures that the parameters are in a convenient range for numerical optimizers and have roughly the same impact on blurriness. That way, common optimizer implementations work with their default parameters. See section ``Data and code availability'' for a link to the source code and examples.

The test dataset presented in Figures \ref{fig:miscalibrated} - \ref{fig:numerical} was obtained using a gold cross-grating replica sample. Data acquisition was conducted on a JEOL GRANDARM2 equipped with a MerlinEM 4R (quad-chip Medipix) system. The physical pixels on the detector chip are 55 µm wide. The scanning coils of the microscope are synchronized with the detector through TTL signals via a BNC cable. The scan array consists of 64 by 64 steps, with each step measuring 12.5 nm. The frames, each comprising 512 by 512 pixels, were each exposed for 1ms, culminating in a total acquisition time of approximately 4 seconds.

The Supplementary Material \citep{Zenodo_supp} contains a full screen capture video with explanation of the very first live adjustment using this method. The test specimen was a semiconductor heterostructure in the bump-bonding region of a PCB, where a random area was selected for the preparation of a TEM lamella using a focused ion beam (FIB) instrument. A prototype of the DECTRIS ARINA detector mounted on a probe corrected JEOL JEM-ARM200F transmission electron microscope was used to record the live data. The microscope was operated at 200kV acceleration voltage in \ac{STEM} mode with a probe of 28.2 mrad semi-convergence angle and 116 pA beam current. The ARINA detector was operated at 50 kHz frame rate with (2x2) binning which resulted in 96x96 detector pixels (detector pixel size: 100 um). This was performed using a prototype implementation of the method described here. Most notably, it didn't use TEMGYM and accelerated raytracing yet, but a preliminary manual ray tracing implementation of the model described above. The detector data was not recorded due to the ephemeral nature of live processing.

\section{Results}

A reliable manual alignment could be found in less than 15 min on the very first attempt with this approach using a prototype implementation. See the Supplementary Material at \citep{Zenodo_supp} for a screen capture video of this first attempt with explanations, and a detailed description. Note how live processing allowed adjustment of microscope parameters and specimen movement interactively. Observing specimen movement allowed rough calibration of scale, rotation and handedness even on a detector with a low number of pixels such as the DECTRIS ARINA, which can only show a small region of interest at coarse spatial resolution. We also note that the high frame rate of the ARINA detector gave a sufficient scan repetition rate to observe the effect of changes in near real time. The rotation and handedness parameters obtained with the digital twin were validated to correspond to the parameters of LiberTEM's center of mass (CoM) analysis \citep{LiberTEM-CoM} and the \ac{SSB} implementation in the Ptychography 4.0 project~\citep{Weber2021}. This enabled successful ptychographic reconstruction using this \ac{SSB} implementation in unrelated experiments following the adjustment in the screen capture video. The results for Overfocus, Scan Pixel Size, Camera Length and Detector Pixel Size have not been validated experimentally yet.

\begin{figure}
    \centering
    \includegraphics[width=230pt]{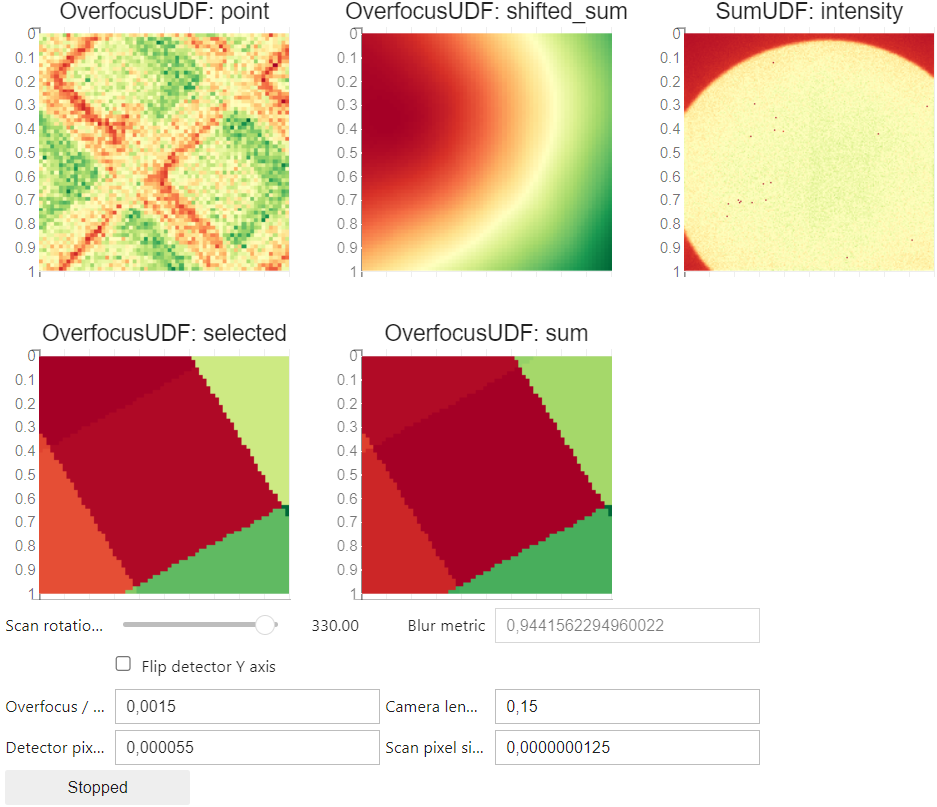}
    \caption{Starting condition with scale and alignment miscalibrated. Control interfaces for the calibration parameters are shown at the bottom. The field ``Blur metric'' shows the blurriness of the ``OverfocusUDF: shifted\_sum'' image as calculated by the algorithm by \cite{Crete2007}. The plot ``OverfocusUDF: point'' shows the trace of a single detector pixel as a function of scan position. ``OverfocusUDF: shifted\_sum'' is the main adjustment plot that shows all detector images superimposed in specimen coordinates according to the transformation calculated by TEMGYM that is derived from the input parameters below the plots. If the transformation by TEMGYM matches the actual transformation by the microscope, this plot is a sharp image of the specimen. Note that it is blurred here and doesn't show any specimen features, meaning the parameters are not correct yet. ``SumUDF: intensity'' shows a sum of all untransformed diffraction patterns. The plot ``OverfocusUDF: selected'' shows the diffraction pattern at the center of the scan grid rotated and scaled to scan coordinates according to the current settings. Structures of the specimen are only recognizable in ``OverfocusUDF: point'', while ``OverfocusUDF: selected'' magnifies the frame so much with the current settings that only a few pixels are in the field of view. The sharp image of the beam-forming aperture shows that the pivot point is adjusted correctly since the aperture is at the same position for each scan point. ``OverfocusUDF: sum'' shows the sum of partially transformed diffraction patterns where only ``Scan Rotation'' and ``Flip Y'' are applied, in contrast to ``SumUDF: intensity'' that shows the sum of the untransformed diffraction patterns where the position of the beam on the detector can be judged.}\label{fig:miscalibrated}
\end{figure}

\begin{figure}
    \centering
    \includegraphics[width=230pt]{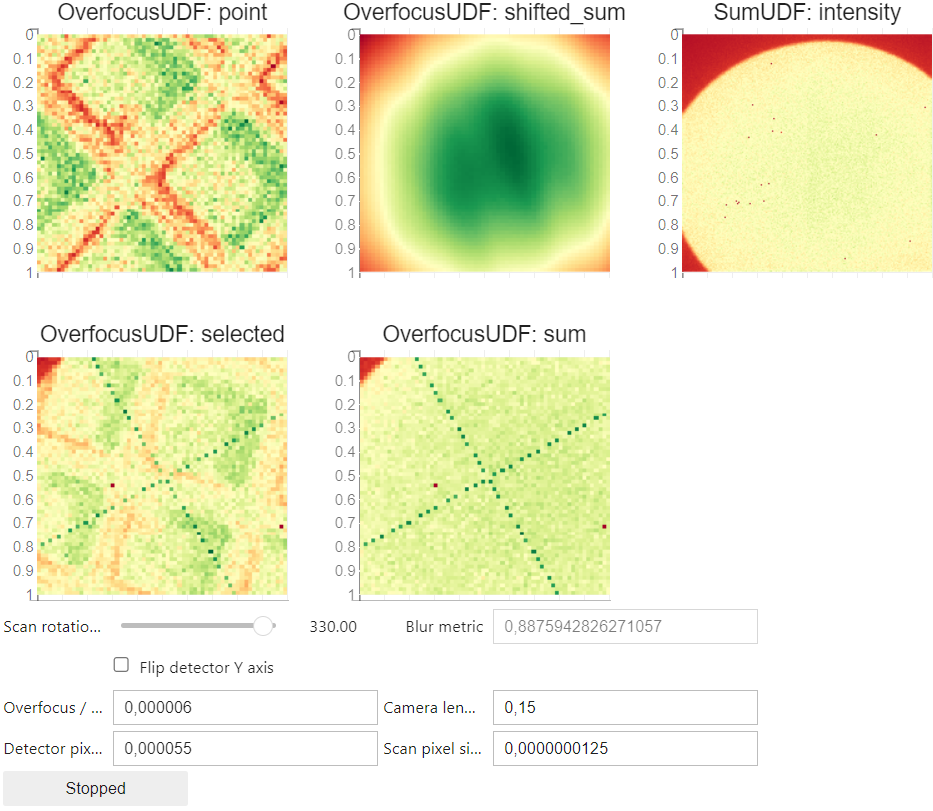}
    \caption{Compared to Figure~\ref{fig:miscalibrated}, the ``Overfocus'' parameter was adjusted so that ``OverfocusUDF: point'' and ``OverfocusUDF: selected'' have approximately the same scale.}\label{fig:scale}
\end{figure}

\begin{figure}
    \centering
    \includegraphics[width=230pt]{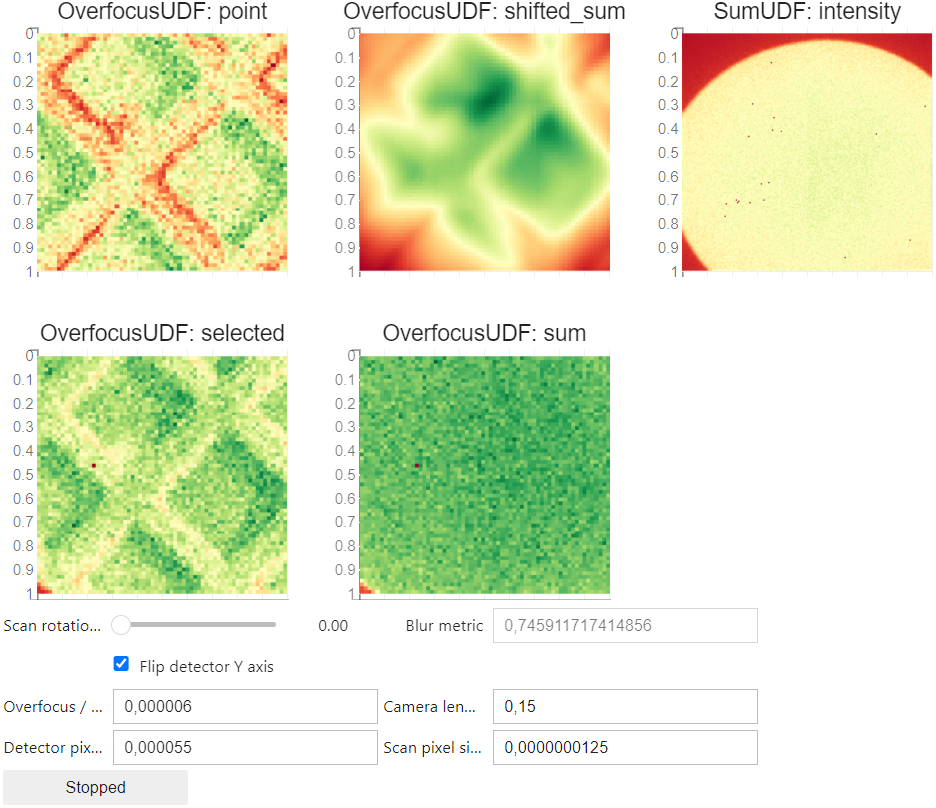}
    \caption{Compared to Figure~\ref{fig:scale}, the ``Scan Rotation'' and ``Flip Y'' parameters were adjusted so that ``OverfocusUDF: point'' and ``OverfocusUDF: selected'' show approximately the same image. Blurred specimen features are starting to appear in ``OverfocusUDF: shifted\_sum''.}\label{fig:rough}
\end{figure}

\begin{figure}
    \centering
    \includegraphics[width=230pt]{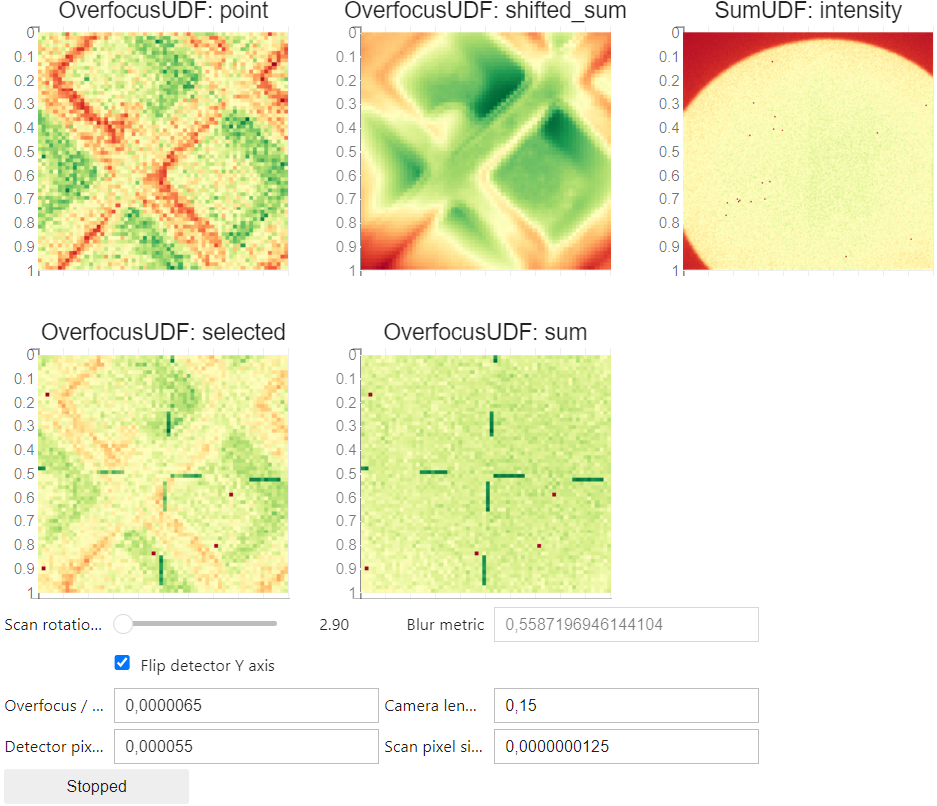}
    \caption{Compared to Figure~\ref{fig:rough}, the ``Overfocus'' and ``Scan Rotation'' parameters are fine-tuned so that ``OverfocusUDF: shifted\_sum'' appears sharp. Note how a small change in ``Overfocus'' and ``Scan Rotation'' compared to Figure~\ref{fig:rough} causes a large difference in ``OverfocusUDF: shifted\_sum'', demonstrating the sensitivity of this method.}\label{fig:fine}
\end{figure}

The Figures~\ref{fig:miscalibrated}-\ref{fig:fine} show the manual adjustment procedure with the test dataset described in the previous section using the most recent software version. The dataset is available at \citep{Zenodo_supp}.

\begin{figure}
    \centering
    \includegraphics[width=230pt]{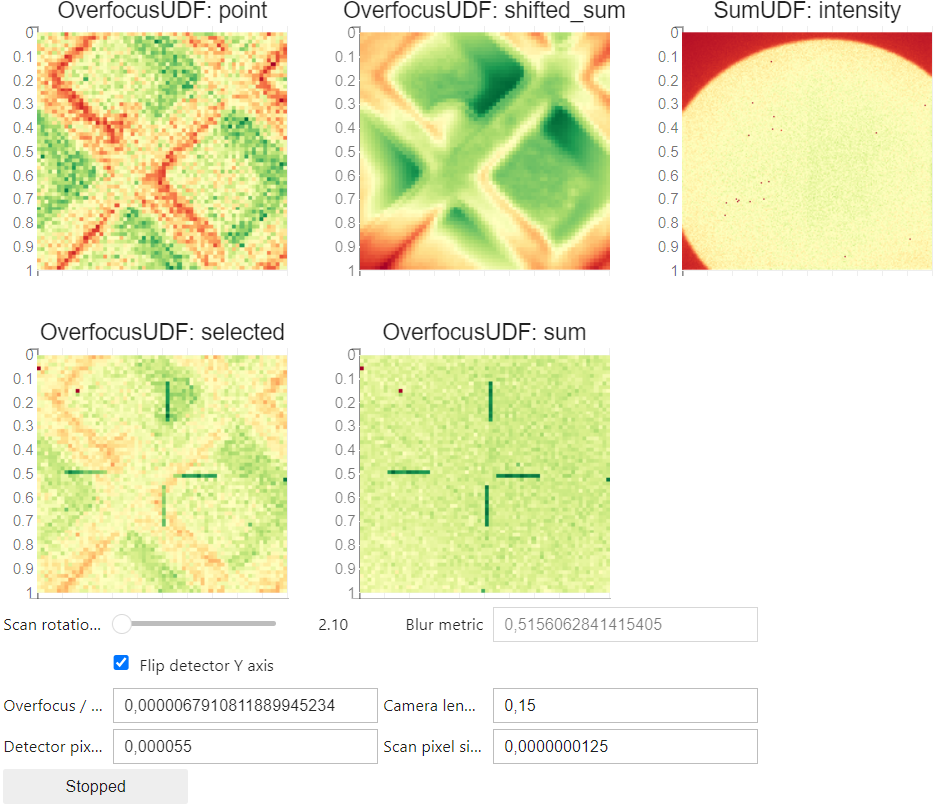}
    \caption{Compared to Figure~\ref{fig:fine}, the ``Overfocus'' and ``Scan Rotation'' parameters are numerically optimized by minimizing the blur metric using global optimization within an interval around the values in Figure~\ref{fig:fine}.}\label{fig:numerical}
\end{figure}

The processing takes 1.5\,s for a single pass over the test file used in Figures~\ref{fig:miscalibrated}-\ref{fig:numerical}, a MIB 4D \ac{STEM} dataset of size 64 $\times$ 64 $\times$ 512 $\times$ 512 with 16 bits per pixel, corresponding to a data rate of about 1.4\,GB/s or 2730 frames per second, on an AMD EPYC 7F72 with 24 physical cores and 256 GB RAM using LiberTEM 0.13 and the default Dask executor. Processing takes 1.38\,s, corresponding to an uncompressed data rate of about 3.5\,GB/s or 47,000 frames per second, for a file of size 256 $\times$ 256 $\times$ 192 $\times$ 192 with 16 bits per pixel in the compressed HDF5 format generated by the DECTRIS ARINA. This is fast enough to follow typical 4D \ac{STEM} data rates live. Live processing allows adjustment of both microscope parameters, such as specimen region, magnification and focus, and model parameters. However, many repeated scans may degrade a specimen through radiation damage or contamination.

A loss function using the blurriness metric \citep{Crete2007} as implemented in the Scikit-Image \citep{Walt2014} function {\tt skimage.measure.blur\_effect} was created using the helper function described in the previous section. It was minimized numerically using the ``simplicial homology global optimization'' (SHGO) algorithm \citep{Endres2018} with ``constrained optimization by linear approximation'' (COBYLA,~\cite{Powell1994}) for local search, as implemented in the SciPy~\citep{Virtanen2020} function {\tt scipy.optimize.shgo}. The result is shown in Figure~\ref{fig:numerical}. The fast processing of LiberTEM and the OverfocusUDF allows numerical optimization within about 12\,s from Figure~\ref{fig:fine} to Figure~\ref{fig:numerical} on the system described above. The optimization converged reliably to near optimal values from various starting values, provided the optimum was within the optimization bounds around the starting value as described in the previous section. The optimization was implemented to update the plots live, which allows observing the convergence behavior and checking if the optimization result is sensible.

The following workflow turned out to be practical:

\begin{enumerate}
    \item Select a specimen region, scan pixel size and other microscope parameters that show good contrast and recognizable features in the ``OverfocusUDF: point'' plot. The method only works at low or medium magnification, not at atomic resolution, since it doesn't take diffraction into account.
    \item Set the Scan Pixel Size, Camera Length and Detector Pixel Size to their correct values, as far as known.
    \item Adjust the focus of the microscope so that specimen features are recognizable on the detector. Ideally, the features on the detector have roughly the same scale as in the ``OverfocusUDF: point'' plot.
    \item Set the Overfocus value according to the overfocus given by the microscope. Specimen features should be recognizable in both the ``OverfocusUDF: point'' and ``OverfocusUDF: selected'' plots now.
    \item If possible, adjust Overfocus so that features in ``OverfocusUDF: point'' and ``OverfocusUDF: selected'' have roughly the same scale. This works well on grids, for example.
    \item Adjust Scan Rotation and Flip Y so that movements of the specimen or low-symmetry specimen features are consistent between ``OverfocusUDF: point'' and ``OverfocusUDF: selected''. A blurred image of the specimen should appear in ``OverfocusUDF: shifted\_sum'' now.
    \item Confirm that the scale is approximately the same between ``OverfocusUDF: point'' and ``OverfocusUDF: selected''.
    \item Fine-tune Scan Rotation and Overfocus until ``OverfocusUDF: shifted\_sum'' is sharp. Confirm that Flip Y is set correctly if the image remains blurred.
    \item Optionally, refine with numerical optimization. This works once ``OverfocusUDF: shifted\_sum'' appears only moderately blurred, meaning the starting values are already close enough.
\end{enumerate}

\section{Discussion}

Live processing was helpful for finding a suitable specimen region and microscope parameters as well as coarse adjustment of Scan Rotation and Flip Y, since observing the effect of specimen movement interactively helps to orient the user. For fine adjustments it is advantageous to record a dataset after coarse adjustment of microscope and alignment parameters that show recognizable features in the superposition image. The user can then continue to adjust on such an offline dataset. LiberTEM allows using offline and live data interchangeably, which helps the implementation of such a workflow. This avoids excessive exposure of the specimen. Furthermore, recording such an overfocused calibration dataset before real measurements can allow later validation of the parameters for data interpretation.

Currently, the implementation uses TEMGYM Basic for the digital twin, i.e.\ a matrix optics approximation. However, the method is not limited to this: The ray tracing engine is exchangeable as long as it accepts the same parameters, meaning a more advanced model such as TEMGYM Advanced~\citep{Landers2023a,Landers2023b} could be used. This software traces electron trajectories through electromagnetic fields and enables the inclusion of 3\textsuperscript{rd} order and higher aberration effects from the objective lens. If the internal interpolation for quick mapping of detector frames is adapted to allow non-linear relations between detector pixel, scan position and specimen pixel, this approach could also be extended to take aberrations into account. Furthermore, the digital twin could be extended to match each optical element of the electron microscope exactly, as already demonstrated in~\cite{Landers2023}. 

The method provides a clear validation for parameters that leaves little room for error or interpretation: If the resulting image is sharp, the parameters of microscope and digital twin are consistent. It can be performed as part of a 4D \ac{STEM} acquisition workflow without switching detector or scan engine. Furthermore, it is sufficiently fast to perform many checks during a session, for example after a change of microscope parameters. In particular if a good initial guess is available, fine adjustment is quick and intuitive, and can even be performed automatically. Since the image used for fine calibration is composed of many detector frames, it has a better signal to noise ratio than individual detector frames or the plot of the central pixel, allowing quick fine calibration at a higher scan rate and low beam current if a good starting value is known.

The current user interface in a Jupyter notebook works as a proof of concept. In the future, a stand-alone tool could be developed or the method could be integrated into the microscope control software. In particular, the actual microscope parameters could be read out digitally and applied to the digital twin. If microscope calibration values instead of model parameters are optimized, it would allow the microscope to operate directly in units that correspond to the digital twin.

The approach is also extensible to other parameters of the microscope that change the detector image in a systematic way and allow superposition of transformed detector images to form a sharp image of an object in the beam, such as a specimen or aperture. As a disadvantage, it only works with 4D \ac{STEM} detectors, not conventional or segmented \ac{STEM} detectors.

It should be investigated if changing the focus strongly in real microscopes can affect the observed rotation or other parameters.

In this paper we already show how parameters can be transformed between three different models or digital twins that describe the projection performed by the microscope: TEMGYM Basic uses ray tracing through a simplified \ac{STEM} model based on a linear approximation for an objective lens and double beam deflectors (Figure~\ref{fig:visualization}). The parameters exposed towards the user (controls in Figure~\ref{fig:miscalibrated}-\ref{fig:numerical}) describe a simpler model where the beam is not deflected by double deflectors or focused by a lens, but where the specimen coordinate system is simply shifted and a focus position is specified (Figure~\ref{fig:abstraction}). Internally, the ``OverfocusUDF'' uses a simple ray transfer matrix that directly performs a mapping from detector coordinates and scan position to specimen pixel coordinates for optimal performance.

Since each model is linear, and provides a good approximation of the actual instrument's behaviour under the chosen conditions, these are equivalent descriptions of the same transformation that are specified in different ways to match the context in which they are used. It also allows matching them exactly using computational methods. The user interface aids the user in finding approximate parameters that match the actual projection by the instrument. Numerical optimization can then be used to fine-tune this result further. In effect, this determines the instrument parameters in a reliable and precise way.

Currently, the commonly used implicit models and parameters for metadata schemas and data analysis only approximate the action of the instrument in specific optical modes such as transmission electron microscopy (TEM), \ac{STEM}, electron energy loss spectroscopy (EELS), etc. The model in Figure~\ref{fig:abstraction} can describe strongly defocused \ac{STEM}, but can't model the projection for in-focus \ac{STEM} since it doesn't consider diffraction, for example. If a more comprehensive or even full physical model is used to describe the projection by the microscope, the approach described here can allow transformation of physical parameters, such as objective lens current, into parameters for any simplified abstract model, such as the \ac{STEM} model here. It also allows transformation of parameters between different models, for example to determine the parameters on a different microscope to perform an equivalent projection. A set of comprehensive model, parameters and calibration can describe the action of the instrument in every detail and in every state in a reproducible way, provided the model is sophisticated enough. At the same time, corresponding parameters for any simplified model can be derived using, for example, a numerical solver. That allows to use the same metadata schema and model framework for any analysis performed with any TEM. That means a comprehensive digital twin constitutes a potential foundation for universal metadata for electron microscopy.

\section{Code and data availability}

The source code for the adjustment code as well as the Jupyter notebook to run it on offline data can be found at \url{https://github.com/LiberTEM/Microscope-Calibration/tree/37066766ab446153ec2ebb3370cb879905cc410b}.

The TEMGYM Basic version used here is \url{https://github.com/TemGym/TemGym/tree/bc28049225489b6ca75bc3377a4c64f53d77eb31}.

The dataset used in the figures for offline calibration, a screen capture video and additional explanation, as well as a runnable Apptainer image with the complete software stack for reproducibility can be found at \url{https://doi.org/10.5281/zenodo.10418769}.

\section{Acronyms}

\begin{acronym}[overfocus]
    \acro{STEM}{scanning transmission electron microscopy}
    \acro{DPC}{differential phase contrast}
    \acro{CoM}{center of mass}
    \acro{SSB}{single side band}
    \acro{UDF}{LiberTEM~\citep{Clausen2023} user-defined function}
\end{acronym}

\section{Financial support}

Swiss National Science Foundation is kindly acknowledged for co-funding the
electron microscope (R'Equip Project 206021\_177020) used herein.

We gratefully acknowledge support by the AIDAS project.

This project has received funding from the European Union's Horizon 2020 research and innovation programme under grant agreement No 823717 -- ESTEEM3.

We gratefully acknowledge support by DECTRIS for providing access to a prototype of the ARINA detector at PSI.

\section{Competing interests}

The authors declare no competing interests.

\end{document}